\documentstyle[12pt]{article}   % PREPRINT %
\def\be{\begin{equation}}
\def\ee{\end{equation}}
\def\beq{\begin{eqnarray}}
\def\eeq{\end{eqnarray}}
\def\n{\nonumber}
\def\bay{\begin{array}}
\def\eay{\end{array}}

\begin{document}
\begin{titlepage}

\title{Conformal Killing Vectors in Spherically Symmetric, Inhomogeneous,
Shear-free, Separable Metric Spacetimes}

\author{ {\footnotesize S. M. Wagh\thanks{E-mail: ciri@bom2.vsnl.net.in},\, R. V.
Saraykar\thanks{E-mail: sarayaka@nagpur.dot.net.in -- Also at
Department of Mathematics, Nagpur University, Nagpur.},\, P. S.
Muktibodh\thanks {E-mail: muktibodh@satyam.net.in --  Also at
Hislop College, Temple Road, Nagpur.}}
\\
\\
{\footnotesize Central India Research Institute, Post Box 606,
Laxminagar, Nagpur 440 022, India}
\\
\\
{\footnotesize and }
\\
{\footnotesize K. S. Govinder\thanks{E-mail: govinder@nu.ac.za} }
\\
\\
{\footnotesize School of Mathematical and Statistical Sciences,
University of Natal, Durban 4041, South Africa}
\\
\date{December 10, 2001} }
\maketitle

\begin{abstract}
In this paper, we find all the Conformal Killing Vectors (CKVs)
and their Lie Algebra for the recently reported \cite{cqg1}
spherically symmetric, shear-free, separable metric spacetimes
with non-vanishing energy or heat flux. We also solve the geodesic
equations of motion for the spacetime under consideration.
\end{abstract}

%\noindent {\em Keywords:} spherically symmetric spacetimes,
%shear--free, energy flux, equation of state, conformal killing
%vectors, Lie Algebra

%\noindent{\em Running head:} Conformal killing vectors in
%Spherically Symmetric ... \vspace{1in}

\centerline{To be submitted: Classical and Quantum Gravity}
\end{titlepage}
%%%%%%%%%%%%%%%%%%%%%%%%%%%%%%%%%%%%%%%%%%%%%%%%%%%%%%%%%%%%%%%%%%%%%

\section{Introduction}
In a recent paper  \cite{cqg1}, all the spherically symmetric,
inhomogeneous, shear-free, separable-metric spacetimes, that could
admit matter with an equation of state $p\,=\,\alpha\rho$ where
$p$ is the pressure, $\rho$ is the density of matter in the
spacetime and $\alpha$ is a constant, were reported. These
spacetimes could be used to model a shear-free spherical star
\cite{censor} or the shear-free universe at large
\cite{smwinhcos}. It is of some definite interest, and, therefore,
it is the purpose of this paper, to investigate the Conformal
Killing Vectors (CKVs) of these spacetimes. We also solve the
corresponding geodesic equations.

The organization of this paper is as follows. In \S \ref{ssissm},
we provide the salient features of the spacetimes of \cite{cqg1}.
In \S \ref{ckvs}, we obtain all the CKVs for these spacetimes. In
particular, we obtain the Killing vectors in \S \ref{killvec}; the
Homothetic Killing vectors in \S \ref{hkillvec} and the
Non-Special Conformal Killing vectors in \S \ref{nospekill}. There
are no Special Conformal Killing vectors for these spacetimes as
pointed out in \S \ref{spekill}. A summary of these results is
provided in \S \ref{sumup} along with the table for conformal
factors for each CKV. Also provided in this section are the
structure constants of the Lie Algebra of the CKVs. In \S
\ref{geodesics}, we solve the geodesic equations of motion for the
spacetime under consideration. In the concluding section, \S
\ref{conclude}, we discuss the physical applications.

\section{The Spacetime} \label{ssissm}
The metric of all the spherically symmetric, inhomogeneous,
shear-free, separable-metric spacetimes, that admit matter with an
equation of state $p\,=\,\alpha\rho$ where $p$ is the pressure,
$\rho$ is the density of matter in the spacetime and $\alpha$ is a
constant, is : \be
ds^2\;=\;-\,y^2(r)\,dt^2\;+\;2\,R^2(t)\,\left(\frac{dy}{dr}\right)
^2\,dr^2\;+\;y^2(r)\,R^2(t) \left[ \,
d\theta^2\;+\;\sin^2{\theta}\,d\phi^2 \,\right] \label{met}\ee It
should be noted that the metric (\ref{met}), in general, has
\cite{cqg1}: $g_{tt}\,=\,-\,y^2\,A^2(t)$. However, a redefinition
of the time coordinate can be made to absorb the temporal function
$A(t)$ without affecting any physical or geometrical properties.

In what follows, we shall denote a time derivative by an overhead
dot and derivative with respect to the radial coordinate by an
overhead prime. Here $(t, r, \theta, \phi)$ are the co-moving
coordinates. In a Lorentz frame, this metric has the Einstein
tensor\beq
G_{tt} &=& \frac{1}{2y^2R^2}\;+\;\frac{3\dot{R}^2}{y^2R^2} \\
G_{tr} &=&
\frac{\sqrt{2}\dot{R}}{y^2R^2} \\
G_{rr}\;=\;G_{\theta\theta}\;=\;G_{\phi\phi}\;&=&
\frac{1}{2y^2R^2}\;-\;\frac{2\ddot{R}}{y^2R}\;-\;\frac{\dot{R}^2}{y^2R^2}
\eeq Note that the $G_{tr}$ component of the Einstein tensor is
non-vanishing for $\dot{R}\,\neq\,0$. The presence of the $G_{tr}$
component of the Einstein tensor implies the presence of an energy
flux across $r\,=\,{\rm constant}$ surfaces in the spacetime of
metric (\ref{met}). This metric admits the following invariants:
\begin{eqnarray}
R &=& \frac{1 - 6 \dot{R}^2 - 6 R \ddot{R}}{R^2y^2} \label{ricci} \\
R_{ab}R^{ab} &=& \frac{1 + 12 \dot{R}^4 - 6 R \ddot{R} +12 R^2 \ddot{R}^2 - 4 \dot{R}^2
(1-3R\ddot{R})}{R^4y^4} \label{contricci} \\
R_{abcd}R^{abcd} &=& \frac{3-4\dot{R}^2+12\dot{R}^4-8R\ddot{R} +
12 R^2\ddot{R}^2}{R^4y^4} \label{kretschmann}
\end{eqnarray} Clearly (\ref{met}) is singular iff $y(r)=0$ for
some $r$ and/or $R(t)=0$ for some $t$.

The fluid four-velocity $U^a$ is co-moving and is given by \be U^a
= \frac{1}{y}\delta^{a}_0 \label{fvelo} \ee The fluid
four-acceleration $\dot{U}_a = U_{a;b}U^b$ is given by  \be
\dot{U}_a = (0, {{y^{\prime}}\over y},0, 0) \label{facc} \ee where
a semicolon denotes a covariant derivative. The spacetime under
consideration has non-zero acceleration unless $y'\,=\,0$.

The presence of the energy flux implies\footnote{While a
non-vanishing radial heat flow is one possible interpretation, it
is not unique.  All one really knows is that there must be radial
energy transport.  This could be caused by heat, or could just as
well be caused by the bulk motion of matter or radiation.} the
energy-momentum tensor of the matter fluid to be of the form \be
T_{ab} = (\rho + p)U_a U_b + p g_{ab} + Q_a U_b + Q_b U_a
\label{u6} \ee where $p$ is the pressure, $\rho$ is the density of
the fluid and $Q^a = ( 0, Q, 0, 0)$ is the radial energy or the
radial heat flux four-vector. Note that both shear and rotation
vanish for the metric (\ref{met}).

\goodbreak
The expansion, $\Theta$, energy or heat flux, $Q$, density, $\rho$
and pressure, $p$, are given by
\beq \Theta &=&\;{U^a}_{;a}\;=\; \frac{3}{y}\,\frac{\dot{R}}{R} \label{expansion}\\ \n \\
Q &=& \frac{-\,1}{y^2\,y^{\prime}}\,\frac{\dot{R}}{R^3} \label{heatflux} \\ \n \\
\rho &=&\frac{1}{y^2\,R^2}\,\left[ \frac{1}{2} \;+\;3\,\dot{R}^2
\label{density} \right] \\ \n \\
p &=&
\frac{1}{2y^2R^2}\,-\,\frac{\dot{R}^2}{y^2R^2}\,-\,\frac{2\ddot{R}}{y^2R}
\label{pressure} \eeq From eqs. (\ref{density}) and
(\ref{pressure}), we obtain \be
\ddot{R}\;=\;\frac{y^2R}{6}\,\left[ \frac{2}{y^2R^2}\;-\;\left(
\rho\,+\,3\,p \right) \right] \label{deceleration} \ee

Therefore, the equation of state, that is, a relation between $p$
and $\rho$, of the matter in the spacetime of (\ref{met})
determines the dynamics - the behavior of the temporal function
$R(t)$ - of the spacetime. The temporal behavior of the heat flux
is also determined by the temporal function $R(t)$.

The spatial or radial nature of the heat flux is determined
primarily by the sign of the quantity $-\,\dot{R}/y'$. The heat
flux is positive, that is, heat flows from lower values of $r$ to
higher values of $r$, when $y^{\prime}$ and $\dot{R}$ have
opposite signs. In the case of these two quantities having the
same sign, the heat flux is negative, that is, heat flows from
higher values of $r$ to lower values of $r$. On the other hand,
the density is, for $y'\,>\,0$, a decreasing function of $r$
corresponding to a region over-dense at its center and, for
$y'\,<\,0$, an increasing function of $r$ corresponding to a
region under-dense at its center.

For $\dot{R}\,>\,0$ ($\dot{R}\,<\,0$), the spacetime under
consideration is expanding (contracting) since the expansion is
positive (negative), $\Theta\,>\,0$ ($\Theta\,<\,0$). For
$\dot{R}\,>\,0$, the heat flux is positive for $y'\,<\,0$ and
negative for $y'\,>\,0$. Therefore, when the spacetime is
expanding, heat flows from under-dense regions to over-dense
regions. For $\dot{R}\,<\,0$, the heat flux is positive for
$y'\,>\,0$ and negative for $y'\,<\,0$. Therefore, when the
spacetime is contracting, heat flows from over-dense regions to
under-dense regions. Any two adjacent regions with opposite signs
for heat flux are joined by heat flow caustics at locations for
which $y'\,=\,0$. Note that the quantity $Q$ blows up at such
locations.

However, we note that the density is finite at locations
$y'\,=\,0$. The density is infinite only at locations
$y(r)\,=\,0$. We also observe that the metric (\ref{met}) is
singular at heat flow caustics {\it i.e.} for $y'\,=\,0$. Such
locations are, however, coordinate singularities since the
curvature invariants (\ref{ricci}), (\ref{contricci}) and
(\ref{kretschmann}) are finite there.

We also observe here that the radial function $y(r)$ is not
determined by the field equations. Therefore, radial attributes of
matter are {\em arbitrary}, meaning, unspecified, for the metric
(\ref{met}). This is in the manner of concentric spheres with each
sphere allowed to possess any value of density, for example. This
is the maximal freedom compatible with the assumption of spherical
symmetry, we may note.

The point $r\,=\,0$ will possess a locally flat neighborhood when
${y'|}_{r\,\sim\,0}\;\approx\;1/\sqrt{2}$. This condition must be
imposed on any $y(r)$. Apart from this condition, the function
$y(r)$ is arbitrary. Other physical considerations, for example
such as those arising from the equation of heat transfer in the
spacetime, could also constrain the function $y(r)$.

A stellar model with heat flux can be obtained for an appropriate
choice of the radial function $y(r)$ \cite{cqg1}. On the other
hand, the metric (\ref{met}) can also provide an inhomogeneous
cosmological model with non-vanishing heat flow.

This essentially completes our description of the spherically
symmetric, inhomogeneous, shear-free, separable-metric spacetimes
that could admit an equation of state $p\,=\,\alpha\,\rho$ for the
matter in the spacetime \cite{cqg1}.

\section{Conformal Killing Vectors}\label{ckvs}
Conformal symmetries are of some importance in the understanding
of spacetime geometry. The conformal motions preserve the angles
between vectors and, hence, the light cone structure of the
spacetime. These symmetries thereby help solve the geodesic
equations of motion for the spacetime under consideration.

A Conformal Killing Vector (CKV) ${\bf X}$ satisfies \be {\cal
L}_{\bf X}\,g_{ij}\;=\;2\,\Phi(x^k)\,g_{ij} \label{ckveq} \ee
where $\Phi(x^k)$ is the conformal factor and $g_{ij}$ is the
spacetime metric tensor. There typically arise the following four
special cases of CKVs, namely,

\begin{tabular}{llll}
1] &Killing vectors &\hspace{.4in}$K_J$ &\hspace{.4in}$\Phi = 0$
\\ \\
2] &Homothetic Killing vectors  &\hspace{.4in}$H_J$
&\hspace{.4in}$\Phi_{,j}=0\neq\Phi$ \\ \\
3] &Special Conformal Killing vectors  &\hspace{.4in}$S_J$
&\hspace{.4in}$\Phi_{,ij}=0\neq\Phi_{,j}\neq\Phi$ \\ \\ 4]
&Non-Special Conformal Killing vectors &\hspace{.4in}$N_J$
&\hspace{.4in}$\Phi_{,ij}\neq0\neq\Phi_{,j}\neq\Phi$
\end{tabular}

\noindent These vectors are of physical significance as they help
produce first integrals. In particular, the Killing vectors
generate the constants of motion and the homothetic Killing
vectors scale distances by the same constant factor and, hence,
preserve the null geodesic affine parameters. Conformal Killing
vectors generate constants of motion along null geodesics. For
details, see \cite{ladies}, \cite{selvan}.

The set of Conformal Killing Vectors forms a Lie Algebra $G_r$ ($r
\leq 15$) with the basis $\{ X_l \}$: ${\cal L}_{
X_l}\,g_{ij}\;=\;2\,\Phi_{l}\,g_{ij}$ such that \be \left[ {X_l,
\, X_j} \right] \;=\;C^k_{lj}{X_k} \hspace{1.5in} {X_l}\Phi_{
j}\,-\,{X_j}\Phi_{l}\;=\;C^k_{lj}\Phi_{k} \ee

The eq. (\ref{ckveq}) for the metric (\ref{met}) reduces to the
system \beq (t,t)
\hspace{.5in} &\frac{y'}{y}X^r\;+\;{X^t}_{,t} &= \Phi \label{tt}\\
(t,r) \hspace{.5in}
&-\,y^2\,{X^t}_{,r}\;+\;2(y')^2R^2{X^r}_{,t}&=0 \label{tr}\\
(t,\theta) \hspace{.5in}
&-\,{X^t}_{,\theta}\;+\;R^2{X^{\theta}}_{,t}&=0 \label{ttheta}
\\ (t,\phi) \hspace{.5in}
&-\,{X^t}_{,\phi}\;+\;R^2\sin^2{\theta}{X^{\phi}}_{,t}&=0 \label{tphi} \\
(r,r) \hspace{.5in}
&\frac{\dot{R}}{R}{X^t}\;+\;\frac{y''}{y'}X^r\;+\;{X^r}_{,r}&=\Phi
\label{rr}
\\ (r,\theta) \hspace{.5in}
&y^2\,{X^{\theta}}_{,r}\;+\;2(y')^2{X^r}_{,\theta}&=0 \label{rtheta} \\
(r,\phi) \hspace{.5in}
&y^2\,\sin^2{\theta}\,{X^{\phi}}_{,r}\;+\;2(y')^2{X^r}_{,\phi}&=0
\label{rphi}
\\ (\theta,\theta)
\hspace{.5in}
&\frac{\dot{R}}{R}{X^t}\;+\;\frac{y'}{y}X^r\;+\;{X^{\theta}}_{,\theta}&=\Phi
\label{thetatheta}
\\ (\theta, \phi) \hspace{.5in} &{X^{\theta}}_{,\phi}\;+\;\sin^2{\theta}
{X^{\phi}}_{,\theta}&=0 \label{thetaphi} \\ (\phi,\phi)
\hspace{.5in}
&\frac{\dot{R}}{R}{X^t}\;+\;\frac{y'}{y}X^r\;+\cot{\theta}X^{\theta}
\;+\;{X^{\phi}}_{,\phi}&=\Phi \label{phiphi} \eeq In what follows,
we obtain the Conformal Killing Vectors from the above equations
case by case.

\subsection{Killing vectors} \label{killvec}
As can be easily verified, the metric (\ref{met}) admits the
following three spacelike Killing vectors: \beq  {X_I} &\equiv
K_I &= \left( 0, \,0, \,0, \,1 \right)  \\
{X_{II}} &\equiv K_{II} &=\left( 0, \,0, \,\sin{\phi},
\,\cos{\phi}\cot{\theta} \right)  \\
{X_{III}} &\equiv K_{III} &=\left( 0, \,0, \,\cos{\phi},
\,-\,\sin{\phi}\cot{\theta} \right) \eeq Now, for any timelike
Killing vector $K_{IV}\;=\;\left(\,K^t,\,0,\,0,\,0\,\right)$
orthogonal to $t\,=\,{\rm constant\/}$ hyper-surfaces, eqs.
(\ref{tr}) - (\ref{phiphi}) are identically satisfied. The eq.
(\ref{tt}) then implies \be X_{IV} \;\equiv \;K_{IV}\; =\;\left(
1, \,0, \,0, \,0 \right) \hspace{1.5in} {\rm condition}\, \dot{R}
\,=\,0 \ee Now, for a Killing vector
$K_V\;=\;\left(\,K^t(t),\,K^r(r),\,0,\,0\,\right)$ that is {\em
not\/} hyper-surface orthogonal, eqs. (\ref{tr}), (\ref{ttheta}),
(\ref{tphi}), (\ref{rtheta}), (\ref{rphi}), (\ref{thetaphi}) can
be satisfied. Then, eqs. (\ref{tt}), (\ref{rr}),
(\ref{thetatheta}), (\ref{phiphi}) imply \be {X_{V}} \;\equiv
K_{V}\; =\;\left( \kappa\,R, \,-\,\frac{y}{y'}\,\kappa\,{\cal C},
\,0, \,0 \right) \hspace{1.5in} {\rm condition}\, \ddot{R}\,=\,0
\ee where $\kappa$ is a constant and $\dot{R}\,=\,{\cal C}$.

The Killing vectors $K_I$, $K_{II}$ and $K_{III}$ are always
spacelike, $K_{IV}$ is always timelike and $K_{V}$ can be
spacelike, null or timelike. Any spherically symmetric spacetime
admits $K_I$ - $K_{III}$ as spacelike Killing vectors. The
conformal factor in each of these cases is zero.

We note that the spacetime admitting $K_{IV}$ as a Killing vector
is a {\em static\/} spacetime. The Killing vector $K_V$ is neither
normal nor tangent to $t\,=\,{\rm constant}$ hyper-surfaces. It is
known \cite{rmsdm} that conformal motions generated by conformal
Killing vectors do not, in general, map a fluid flow conformally.

\subsection{Homothetic Killing vectors} \label{hkillvec}
In this case, $\Phi\,=\,{\rm constant}$.

For $H_I\,=\,\left(\,0,\,H^r,\,0,\,0\,\right)$, eqs. (\ref{tr}),
(\ref{ttheta}), (\ref{tphi}), (\ref{rtheta}), (\ref{rphi}),
(\ref{thetaphi}) can be satisfied. Eqs. (\ref{tt}), (\ref{rr}),
(\ref{thetatheta}), (\ref{phiphi}), then, imply \be X_{VI}
\;\equiv\; H_I \;=\; \left( \,0, \,B_I\,\frac{y}{y'}, \,0, \,0
\,\right) \ee For $H_{II}\,=\,\left(\,H^t,\,0,\,0,\,0\,\right)$,
eqs. (\ref{tr}), (\ref{ttheta}), (\ref{tphi}), (\ref{rtheta}),
(\ref{rphi}), (\ref{thetaphi}) can be satisfied. Eqs. (\ref{tt}),
(\ref{rr}), (\ref{thetatheta}), (\ref{phiphi}) yield \be X_{VII}
\;\equiv\; H_{II} \,=\, \left( \,B_{II}\, \frac{R}{\dot{R}}, \,0,
\,0, \,0 \,\right) \hspace{1.in} {\rm condition}\, \ddot{R}\,=\,0
\ee For $H_{III}\,=\,\left(\,H^t,\,0,\,0,\,H^{\phi}\,\right)$,
eqs. (\ref{tr}), (\ref{ttheta}), (\ref{rtheta}), (\ref{rphi}),
(\ref{thetaphi}) can be satisfied. Eqs. (\ref{tt}), (\ref{tphi}),
(\ref{rr}), (\ref{thetatheta}), (\ref{phiphi}) provide us with \be
X_{VIII} \;\equiv\; H_{III}\;=\;\left(
\,B_{III}\,\frac{R}{\dot{R}}, \,0,\,0, \,1 \,\right) \hspace{1.in}
{\rm condition}\, \ddot{R}\,=\,0 \ee For
$H_{IV}\,=\,\left(\,H^t,\,0,\,H^{\theta},\,H^{\phi}\,\right)$,
(\ref{tt}) and (\ref{rr}) fix $H^t$. As for the angular
components, there exist two linearly independent solutions,
namely, $H^{\theta}(\phi)\,=\,\sin{\phi}$ and
$H^{\theta}(\phi)\,=\,\cos{\phi}$. We, therefore, obtain \be
X_{IX} \;\equiv\; H_{IV}\;=\; \left( \,B_{IV}\,\frac{R}{\dot{R}},
\,0, \,\sin{\phi}, \,\cos{\phi}\cot{\theta} \,\right)
\hspace{1.in} {\rm condition}\, \ddot{R}\,=\,0  \ee \be X_{X}
\;\equiv\; H_{V}\;=\;\left( \,B_V\,\frac{R}{\dot{R}} , \,0,
\,\cos{\phi}, \,-\,\sin{\phi}\cot{\theta} \,\right) \hspace{1.in}
{\rm condition}\, \ddot{R}\,=\,0 \ee where $B_I$, $B_{II}$,
$B_{\scriptscriptstyle{III}}$, $B_{IV}$, $B_V$ are constants.

The vector $H_I$ is always spacelike, $H_{II}$ is always timelike
and $H_{III}$, $H_{IV}$ and $H_V$ can be spacelike, null or
timelike depending on the values of corresponding constants. The
conformal factor in each of the above five cases is equal to the
corresponding constant $B_J$, ($\scriptstyle{J\,=\,I,
\,II,\,III,\,IV,\,V}$).

\subsection{Special CKVs} \label{spekill}
There are {\em no\/} Special Conformal Killing vectors for the
metric (\ref{met}). The conditions $\Phi_{,ij}=0$, together with
the equations (\ref{tt}) - (\ref{phiphi}) force $\Phi={\rm
constant}$.

\subsection{Non-special CKVs} \label{nospekill}
For Non-Special Conformal Killing vectors $\Phi_{,ij}\,\neq\,0$.

In this case we find two vectors given by \beq X_{XI} &=\; N_I
&=\; R \, \frac{\partial\ }{\partial t} \\
X_{XII}&=\; N_{II} &=\; R \,\log y \,\frac{\partial\ }{\partial t}
+ \frac{y'}{y}\, \int \frac{d t}{2 R} \frac{\partial\ }{\partial
r} \eeq with conformal factors given by \beq \phi &=& \dot{R} \\
\phi &=& \dot{R}\, \log y + \int \frac{d t}{2 R} \eeq
respectively.

\section{Summary} \label{sumup}
It should be noted that out of the twelve CKVs given above only
the $X_I$, $X_{II}$, $X_{III}$, $X_{VI}$, $X_{XI}$, $X_{XII}$ are
linearly independent. Other conformal Killing vectors reported
earlier are obtainable from these six vectors under the stated
conditions.

In what follows, we summarize the conformal factors for these six
vectors. We also provide in this section the Lie Brackets for
these six vectors.

The conformal factors for these CKVs  are as  follows:
\bigskip

\begin{tabular}{clccc} \hline  \hline \\ \phantom{m}\hspace{.05in}
CKV &$\rightarrow$  & $\hspace{.5in}X_I/K_I$
&$\hspace{.5in}X_{II}/K_{II}$ &$\hspace{.5in}X_{III}/K_{III}$ \\
\\  \hline
\\  \phantom{m}\hspace{.05in} $\Phi$&$ \rightarrow$ &$\hspace{.5in}0$
&$\hspace{.5in}0$ &$\hspace{.5in}0$  \\ \\ \hline \hline \\
\phantom{m}\hspace{.05in} CKV &$\rightarrow$ \hspace{.05in}
&$\hspace{.5in}X_{VI}/H_I$ &$\hspace{.5in}X_{XI}/N_I$ &$\hspace{.5in}X_{XII}/N_{II}$  \\ \\
\hline
\\ \phantom{m}\hspace{.05in} $\Phi$&$ \rightarrow$ \hspace{.05in}
&$\hspace{.5in}B_I$ &$\hspace{.5in}\dot{R}$ &\hspace{.5in}$\dot{R} \log y + \int \frac{d t}{2 R} $ \\ \\
\hline \hline
\end{tabular}

The above CKVs\footnote{Note: The above results have been
confirmed using the differential equation solver within {\tt
PROGRAM LIE} \cite{head}.} form the basis of the Lie Algebra of
CKVs for the metric (\ref{met}). The Lie Algebra of CKVs is
therefore six dimensional, that is, $G_{6}$. The Lie brackets and
corresponding structure constants are easily obtained as follows:
\bigskip

\noindent\begin{eqnarray*} [K_I,K_{II}]&=&K_{III}\\
{}[K_{II},K_{III}] &=& K_I \\
{} [K_{III},K_{I}] &=& K_{II}\\
{}[H_I,N_{II}] &=& N_I \\
{} [N_I, N_{II}] &=& \frac12 H_I
\end{eqnarray*}

We note that the three Killing vectors - $K_I$, $K_{II}$,
$K_{III}$ - form a sub-algebra corresponding to the rotational
symmetry of the space-time. On the other hand, the remaining three
Conformal Killing Vectors - $H_I$, $N_I$, $N_{II}$ - form another
sub-algebra isomorphic to the Lie Algebra corresponding to the
group $E^+(2)$ \cite{miller}. This is realized by setting
$U_1=H_I, U_2=-\,\sqrt{2}\, i\, N_I, U_3 = \sqrt{2}\,i\,N_{II}$.
This is an interesting aspect of the space-time of the metric
(\ref{met}).

\section{Geodesic equations of motion} \label{geodesics}

The Lagrangian for the metric (\ref{met}) is \be 2\,{\cal
L}\;=\;-\,y^2\,\tilde{t}^2\;+\;R^2\,\left[\,
2\,(y')^2\,\tilde{r}^2
\;+\;y^2\,\tilde{\theta}^2\;+\;y^2\,\sin^2{\theta}\,\tilde{\phi}^2
\, \right] \label{lagrange} \ee where an overhead $\,\tilde{}\,$
denotes a derivative with respect to the affine parameter $s$
along the geodesic. Further, ${\cal L}$ takes values $0$ for null,
$+\,1$ for spacelike and $-\,1$ for timelike geodesics.

The geodesic equations of motion are obtained from the
Euler-Lagrange equations: \be \frac{d}{ds}\left(\frac{\partial
(2{\cal L})}{\partial\dot{x}^a}\right)\;=\; \frac{\partial (2{\cal
L})}{\partial x^a} \ee In what follows, we will use
(\ref{lagrange}) to reduce the terms after rearrangement.

The $r$-equation is obtained as: \be \frac{d}{ds} \left(
\,R^2\,y\,y'\,\tilde{r} \,\right)\;=\;{\cal L} \ee and can be
easily integrated to obtain \be \tilde{r}\;=\;\frac{{\cal
L}\,s\,+\,k_1}{R^2\,y\,y'} \label{rgeoeq} \ee

The $\phi$-equation is \be \frac{d}{ds}
\left(\,R^2\,y^2\,\sin^2{\theta}\,\tilde{\phi} \,\right)\;=\;0 \ee
and it implies \be
\tilde{\phi}\;=\;\frac{k_2}{R^2\,y^2\,\sin^2{\theta}}
\label{phigeoeq} \ee where $k$ is a constant of integration.

The $\theta$-equation is obtained as \be \frac{d}{ds} \left(\,
R^2\, y^2\,\tilde{\theta} \,\right)
\;=\;R^2\,y^2\,\sin{\theta}\,\cos{\theta}\,\tilde{\phi}^2
\label{thetageoeq} \ee On using (\ref{phigeoeq}) and rearranging
terms, we obtain \be \frac{d}{ds} \left[
\,\left(\,R^2\,y^2\,\tilde{\theta}\,\right)^2\;+\;k_2^2\cot^2{\theta}
\,\right]\;=\;0 \ee which implies \[
\left(\,R^2\,y^2\,\tilde{\theta}\,\right)^2\;+\;k_2^2\cot^2{\theta}
\;=\;k_3 \] where $k_3$ is an integration constant. Hence, \be
\tilde{\theta}\;=\;\frac{\pm\,1}{R^2\,y^2}\,\sqrt{\,k_3\,-\,k_2^2\,\cot^2{\theta}}
\ee

The $t$-equation is obtainable as \be \frac{d}{ds}\left(\,
R\,y^2\,\tilde{t} \,\right)\;=\;-\,2\,{\cal L}\,\dot{R}
\label{teq} \ee where we have used (\ref{lagrange}) to reduce
terms.

It is difficult to integrate this equation directly. However, the
solution is obtainable from (\ref{lagrange}) using (\ref{rgeoeq}),
(\ref{phigeoeq}), (\ref{thetageoeq}) as \be
\tilde{t}\;=\;\sqrt{-\,\frac{2{\cal
L}}{y^2}\;+\;\frac{1}{y^4R^2}\left[ 2\,({\cal
L}s\,+\,s)^2\,+\,k_3\,+\,k_2^2\right]} \ee Therefore, the geodesic
of the metric (\ref{met}) has the tangent vector \be
T^a\;=\;\left(
\,\tilde{t},\,\tilde{r},\,\tilde{\theta},\,\tilde{\phi}\,\right)
\ee with $\tilde{t}$, $\tilde{r}$, $\tilde{\theta}$,
$\tilde{\phi}$ as obtained above.

Of course, for null geodesics, we obtain by direct integration of
(\ref{teq}): \be Ry^2\tilde{t}\;=\;\delta\ee where $\delta$ is a
constant. Therefore, in general, the null geodesic of the metric
(\ref{met}) has the tangent vector \be
n^a\;=\;\frac{1}{y^2\,R^2}\,\left(\,
\delta\,R,\;\frac{k_1\,y}{y'},\;\pm\,
\sqrt{k_3\;-\;k_2^2\cot^2{\theta}},\;\frac{k_2}{\sin^2{\theta}}
\,\right) \ee with the different constants satisfying a relation
obtainable from $n^a\,n_a\,=\,0$. In particular, for the {\em
radial null geodesic\/} we have \be \delta\;=\;\pm\,\sqrt{2}\,k_1
\ee Along a null geodesic $x^a(s)$ with tangent vector
$n^a\,=\,dx^a/ds$, the following holds: \be
{n^a}_{;\,b}\,n^b\,=\,0\,=\,n^a\,n_a \ee Any conformal Killing
vector, ${\bf X}$, generates along a null geodesic a constant of
the motion: \be \frac{d}{ds} \left( \, {\bf X}\bullet {\bf n}
\,\right)\;=\;{X}_{a;\,b}\,n^a\,n^b\;=
\;2\,\Phi\,g_{ab}\,n^a\,n^b\;=\;0 \ee Each geodesic is specified
by six parameters: four to determine a point on it and two to
determine the direction of its tangent. Thus, the general solution
of the null geodesic equation for the metric (\ref{met}) depends
on six functionally independent constants of null geodesic motion.
These constants of motion are obtainable from the six conformal
Killing vectors and the above tangent to the null geodesic
trivially.

\section{Discussion} \label{conclude}
Most of the known solutions of the Einstein field equations admit
some nontrivial isometry group. Further, there are many solutions
with conformal symmetry, but in most of these the symmetry is
actually homothetic. Much physical insight on astrophysical and
cosmological questions has been obtained from the study of such
solutions. Of course, with the assumption of the spherical
symmetry, there is the rotational isometry group associated with
the spacetime. The question  then is of obtaining further
symmetries of the spherically symmetric spacetime. In this
respect, our spacetime here is interesting since it has only one
homothetic symmetry but two conformal symmetries. A homothetic
Killing vector leads to self-similarity while scaling all the
distances by the same constant factor. The conformal symmetries
provide the constants of geodesic motion along null trajectories
of massless particles.

In this paper, we obtained all the conformal Killing vectors for
the metric (\ref{met}) and their Lie Algebra. We showed that, for
the space-time of the metric (\ref{met}), there exists a
sub-algebra isomorphic to the Lie Algebra of $E^+(2)$. This Lie
Algebra has an associated Local Lie Transformation Group $E^+(2)$
that is a group of distance preserving transformations in the
plane $R_2$.

We also obtained explicitly the tangent vector to the geodesics of
the metric (\ref{met}). In particular, for the trajectories of
massless particles or the null geodesics, the constants of motion
can be obtained from the CKVs presented. The results obtained here
are of importance for the study of astrophysical problems
involving stellar models or for the study of inhomogeneous
cosmological models based on the metric (\ref{met}).

\section*{Acknowledgements}

KSG thanks the University of Natal and the National Research
Foundation for ongoing support.  He also thanks CIRI for their
kind hospitality during the course of this work.

\pagebreak

\end{document}